\documentclass[twocolumn,showpacs,amsmath]{revtex4}
\usepackage{graphicx}%
\usepackage{dcolumn}
\usepackage{amsmath}
\usepackage{here}
\begin{document}
%\DeclareRobustCommand{\baselinestretch{2}}

\title{Soliton transverse instabilities in anisotropic nonlocal self-focusing media}

\author{Kristian Motzek and Friedemann Kaiser}

\affiliation{Institute of Applied Physics, Darmstadt University of
Technology, D-64289 Darmstadt, Germany}

\author{Wen-Hen Chu and Ming-Feng Shih}

\affiliation{Physics Department, National Taiwan University,
Taipei, 106, Taiwan}

\author{Yuri Kivshar}

\affiliation{Nonlinear Physics Group, Research School of Physical
Sciences and Engineering, Australian National University, Canberra
ACT 0200, Australia}

\begin{abstract}
We study, both theoretically and experimentally, the transverse
modulational instability of spatial stripe solitons in anisotropic
nonlocal photorefractive media. We demonstrate that the
instability scenarios depend strongly on the stripe orientation,
but the anisotropy-induced features are largely suppressed for
spatial solitons created by self-trapping of partially incoherent
light.
\end{abstract}

\maketitle

Nonlinearity-driven instabilities have been studied in many
different branches of physics, since they provide simple means to
observe strongly nonlinear effects in Nature. Transverse (or
symmetry-breaking) instabilities of solitary waves have been
predicted theoretically almost 30 years ago~\cite{zakharov}, but
only recently both transverse and spatiotemporal instabilities
were observed for different types of bright and dark spatial
optical solitons~\cite{review,book}.

Many of the experimental studies of spatial optical solitons in a
nonlinear bulk medium are being carried out for photorefractive
crystals known to exhibit an anisotropic nonlocal response
characterized by an asymmetric change of the refractive index.
Given the strong anisotropy of the photorefractive nonlinearities,
it is crucially important whether the theoretical results obtained
mostly for isotropic nonlinear media~\cite{review} can be applied,
at least qualitatively, to the case of anisotropic nonlocal media.
In particular, the transverse instabilities of solitary waves,
that develop under the action of higher-order perturbations or
temporal effects, are known to initiate a breakup of a stripe
soliton, and several different scenarios of the breakup dynamics
are known~\cite{review}. The most interesting scenario is the
generation of new stable localized structures. In the scalar case,
this corresponds to a breakup of a soliton stripe into
(2+1)-dimensional bright solitons in a self-focusing medium.

The purpose of this Letter is twofold. First, we employ a simple
model of an anisotropic nonlocal medium which takes into account
important properties of photorefractive nonlinearities~\cite{zoz}
and demonstrate numerically that the classical scenario of the
soliton transverse instability, namely the stripe breakup and the
formation of two-dimensional spatial solitons~\cite{review},
depends dramatically on the stripe orientation; these results are
fully confirmed by our experimental studies of the breakup of the
vertical, horizontal, and tilted soliton stripes. Second, we
generalize the familiar coherence-function approach, which is
employed for describing the propagation of partially incoherent
light in nonlinear media, and study the transverse instability of
partially incoherent soliton stripes in anisotropic nonlocal
media. We demonstrate theoretically and confirm experimentally
that strong anisotropy-driven features of photorefractive
nonlinearity are largely suppressed by spatial partial incoherence
of light.

%%%%%%%%%%%%%%%% model %%%%%%%%%%%%%%%%%%%%%%%

We consider the propagation of a single optical beam with the
slowly varying amplitude $E$ in a biased photorefractive crystal,
described by the paraxial equation
\begin{equation}
i \frac{\partial E}{\partial z} + \frac{1}{2}\nabla_\perp^2 E
=\frac{\partial \varphi}{\partial x} E, \label{eq1}
\end{equation}
where $\nabla_\perp$ stands for the transverse gradient in the
plane perpendicular to the propagation direction $z$ and $\partial
\varphi / \partial x$ yields the electric field inside the
crystal. We assume that an external electric field is applied to
the crystal, and it is parallel to the direction of $x$, so that
the electric potential $\varphi$ is defined by the potential
equation~\cite{zoz}
\begin{equation}
\nabla_\perp^2
\varphi+\nabla_\perp\varphi\nabla_\perp\ln(1+I)=\frac{\partial}{\partial
x} \ln(1+I), \label{eq2}
\end{equation}
where $I=|E|^2$ is the light intensity inside the crystal.

%%%%%%%%%%%%%%%%%% numerics %%%%%%%%%%%%%%%%%%%%%%%%%%%%

\begin{figure}
\centerline{\includegraphics[width=3.0in]{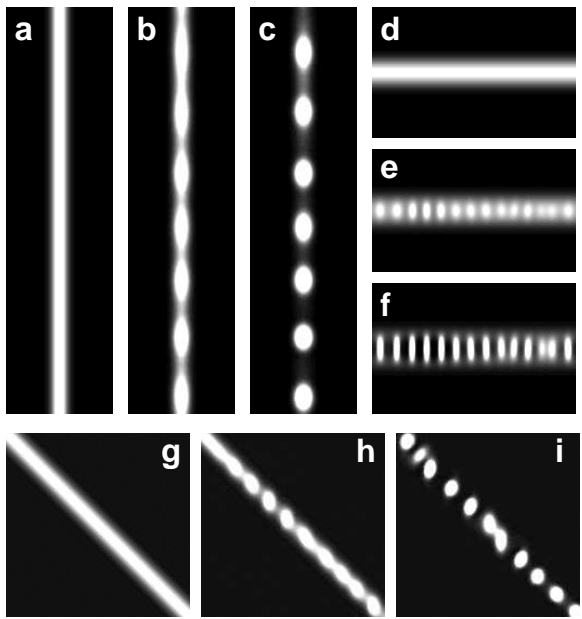}}
\caption{Numerical results for the coherent stripe propagation:
(a-c) the soliton stripe is perpendicular to the external electric
field, (d-f) the soliton stripe is parallel to the field, and
(g-i) an intermediate orientation of a tilted stripe.}
\label{fig_theory}
\end{figure}

First, we study numerically the nonlinear evolution of a narrow
stripe oriented perpendicular to the external electric field.
Figures~\ref{fig_theory}(a-c) show the visualized images of the
nonlinear evolution at the input [(a)] and at two different
propagation distances [(b),(c)]. In numerical simulations, the
initial diameter of the input Gaussian beam was chosen to be close
to that of a solitary solution, so the width of the vertical beam
remains roughly the same. As a result, narrow beams evolve in a
fashion that is very similar to the formation of
quasi-one-dimensional spatial solitons. Increasing nonlinearity
(i.e. the applied field) leads to a breakup of the vertical stripe
due to the transverse modulation instability, first discussed for
photorefractive crystals by Mamaev {\em et al.}~\cite{zoz2,zoz3}.
At the initial stage of the breakup all spatial harmonics of the
noise are small and each is amplified exponentially with its own
growth rate. The fastest-growing modes become noticeable first and
at later stages determine the characteristic spatial scale of the
breakup. At larger propagation distances (larger nonlinearities)
the beam transforms into an array of (2+1)-dimensional bright
solitons [see Fig.~\ref{fig_theory}(c)].

Figures~\ref{fig_theory}(d-f) show the evolution of a stripe
parallel to the external field. Because solitary solutions of this
form do not exist, the stripe initially diffracts. Subsequently, a
growing spatial modulation [(e)] eventually breaks up the stripe
into vertical elongated filaments [(f)] but not solitons.
Figures~\ref{fig_theory}(g-i) show the same effect for the
intermediate orientation of the stripe.

\begin{figure}
\centerline{\includegraphics[width=3.2in]{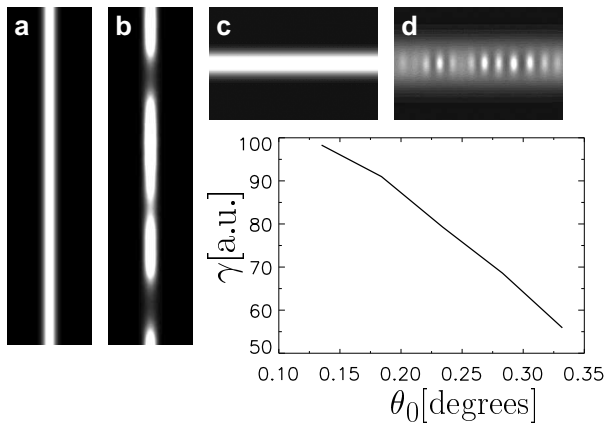}} \caption{
Numerical results for the partially incoherent stripes. (a,b)
Stripe is perpendicular to the external field, and (c,d) stripe is
parallel to the field. Shown are the input [(a,c)] and output
[(b,d)] beams. Insert shows the growth rate of the transverse
instability vs. the coherence parameter $\theta_0$.}
\label{incoh_T}
\end{figure}

\begin{figure}
\centerline{\includegraphics[width=3.1in]{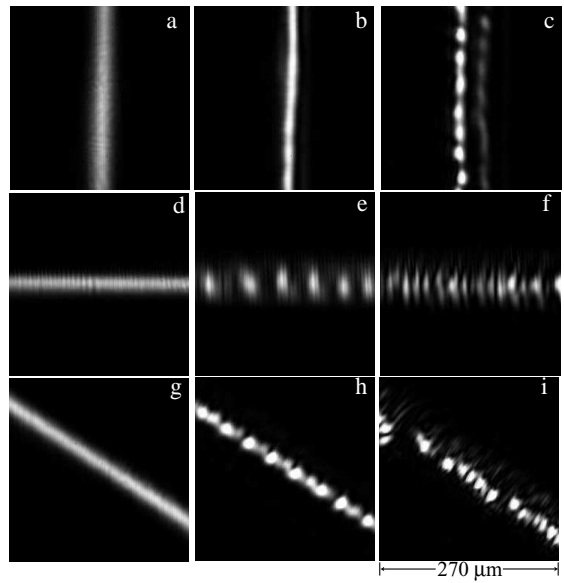}}
\caption{Experimental observation of the transverse instability of
coherent soliton stripe. (a,d,g) inputs for three orientations of
the stripes.  All other images are taken at the crystal output
(length is 7 mm) for the applied voltage of (b,h) 1 kV, (c,i) 2
kV, (e) 0.5 kV, and (f) 1 kV, respectively.} \label{coh}
\end{figure}

%%%%%%%%%%%%%%%%%%% numerics_incoherent $%%%%%%%%%%%%%%%%%

We have also studied numerically the propagation of partially
incoherent light stripes in an anisotropic photorefractive medium
described by the model (\ref{eq1}), (\ref{eq2}). We have extended
the coherent density approach~\cite{IncNum}, earlier used for the
study of isotropic media (see, e.g., Ref.~\cite{inst_exp}), and
described the anisotropic nonlocal response according to Eq.
(\ref{eq2}). The coherent density approach is based on the fact
that partially incoherent light can be described by a
superposition of mutually incoherent light beams that are tilted
with respect to the $z$-axis at different angles. One thus makes
the ansatz that the partially incoherent light stripe consists of
many coherent, but mutually incoherent light stripes $E_j$:
$I=\sum_j |E_j|^2$. By setting $|E_j|^2=G(j\vartheta) I$, where
$G(\theta)=(\pi^{1/2}\theta_0)^{-1}\exp(-\theta^2/\theta_0^2)$ is
the angular power spectrum, one obtains a partially incoherent
light stripe whose coherence is determined by the parameter
$\theta_0$, i.e. less coherence means bigger $\theta_0$. Here,
$j\vartheta$ is the angle at which the $j$-th beam is tilted with
respect to the $z$-axis. If the light is incoherent along the
$y$-axis (vertical stripes), they are tilted in $y$-direction,
whereas they are tilted in the $x$-direction if the light is to be
incoherent along that axis (horizontal stripes).

\begin{figure}
\centerline{\includegraphics[width=3.0in]{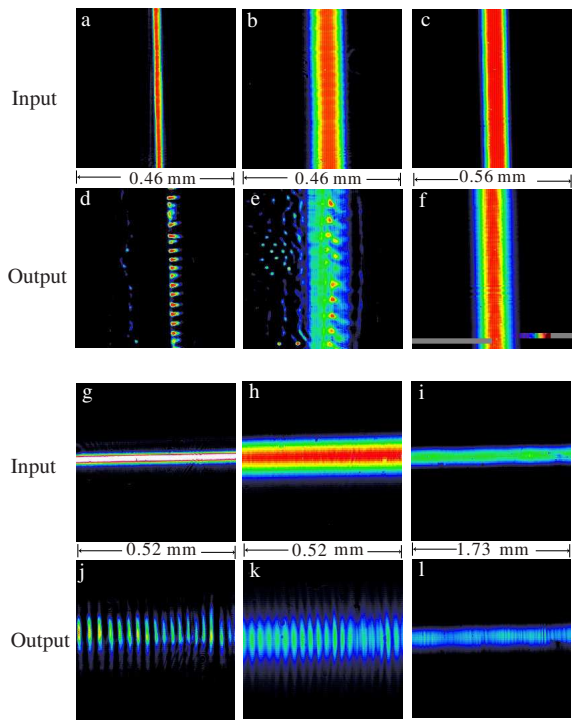}}
\caption{Experimental results for the stripes oriented
perpendicular (two upper rows) and parallel (two lower rows) to
the electric field. Images (a,g) and (d,j) correspond to the input
and output of the coherent light propagation (7 mm), presented for
comparison; whereas all other images correspond to partially
incoherent light.} \label{incoh1}
\end{figure}

Figures~\ref{incoh_T}(a,b)  show our numerical results for the
propagation of a vertical stripe close to the stability threshold
with the degree of incoherence determined by $\theta_0 =
0.43^\circ$. The external electric field is perpendicular to the
stripe. The most obvious difference to the scenario of the
coherent-stripe decay is that the filaments are much more
elongated. Furthermore, they change their profile only very slowly
as they propagate and thus can be considered as incoherent
solitons. Larger values of $\theta_0$ (i.e. $\theta_0 >
0.43^\circ$) correspond to a complete suppression of the soliton
transverse instability.

 Figures~\ref{incoh_T}(c,d) show the propagation of
an incoherent light stripe parallel to the external electric field
with $\theta_0 = 0.56^\circ$. Increasing the incoherence further
leads to the case where the beam diffracts before the transverse
instability can set in. Obviously the filaments have the same size
as in the coherent case, but there is more intensity lost to
radiation. This confirms previous results that it is very
difficult to obtain solitary structures that are elongated along
the axis of the external field.

%%%%%%%%%%%%%%%%%%%%%%%  incoherent experiment %%%%%%%%%%%%%%%%%%%%%%%%

We performed a number of experiments to study the development of
the anisotropy-driven soliton transverse instability in
photorefractive crystals, for both coherent [see Fig.~\ref{coh}]
and partially incoherent [see Fig.~\ref{incoh1}] soliton stripes.
Experiments are conducted in a photorefractive SBN:61 crystal in a
setup similar to that employed for the observation of the
self-trapping of partially incoherent light~\cite{prl_incoh}. We
also illuminate the entire crystal with a background light with
intensity much stronger than that of the soliton stripe to make
the nonlinearity close to a Kerr-type self-focusing nonlinearity.

The beam is made spatially incoherent by passing it through a
rotating diffuser. The rotating diffuser provides a new phase and
amplitude distribution every 1 ms, which is much shorter than the
response time ($\sim$ 1 s) of the medium. We follow Anastassiou
{\em et al.}~\cite{inst_exp} and generate a beam which is very
narrow and fully coherent in one direction (say $x$), yet uniform
and partially incoherent in the other direction (say $y$).

Figures~\ref{coh}(a-c) and \ref{incoh1}(a,d) show the development
of the transverse instability for the coherent stripe
perpendicular to the electric field direction. As the degree of
incoherence grows, the instability weakens [see
Fig.~\ref{incoh1}(b,e)], and then it disappears completely [see
Fig.~\ref{incoh1}(c,f)] because, in order to observe the
instability, the value of the nonlinearity has to exceed a
threshold imposed by the degree of spatial
coherence~\cite{inst_exp}. The nonlinearity is turned on by
applying a voltage of $3 kV$ to the photorefractive crystal with
r$_{33}=250 pm/V$. Figures~\ref{coh}(d-f) and ~\ref{incoh1}(g-l)
show the development of instability for the stripe parallel to the
field. An important observation is that the strong
anisotropy-driven effects observed for coherent light are largely
suppressed when the degree of spatial incoherence exceeds a
threshold [cf. Fig. \ref{coh}(f) and \ref{incoh1}(l)].

%%%%%%%%%%%%%%%%% conclusion %%%%%%%%%%%%%%%%%%%%%%%%

In conclusion, we have studied theoretically and experimentally
the transverse modulational instability of coherent and partially
incoherent soliton stripes in photorefractive crystals. We have
demonstrated a number of novel, anisotropy-driven features of the
stripe instability, as well as analyzed the effect of partial
incoherence.

Y.K. thanks Glen McCarthy for valuable discussions, and the
Physics Department of the Taiwan University for hospitality. This
work was partially supported by the Australian Academy of Science
and the National Science Council, Taiwan.

\end{document}